# 3-SAT Polynomial Solution of Knowledge Recognition Algorithm

Han Xiao Wen    David Zhu    Cuifeng Zhou
PKU Biocity No. 39 Shang Di Xi Lu, Haidian
Beijing, 100085 China

Summary

This paper introduces a **knowledge recognition algorithm** (KRA) for solving the 3SAT problem in polynomial time. KRA learns **member-class relations** and retrieves information iteratively through deductive and reductive reasoning. It applies the principle of Chinese **COVA**[*] (equivalent to a set of eight 3-variable conjunctive clauses) and eliminates the "OR" operation to solve 3-SAT problem. That is, KRA does not search the assignment directly. It recognizes the complements as rejections at each level of the set through **iterative set** relation recognition. KRA recognizes which conjunctive 3-variable-clause is not satisfiable. If all the eight clauses of any set of 3-variable clauses are rejected, then there is not an assignment for the formula. If there is at least one clause in each set that remains, then there is at least one assignment that is the union of clauses of each set. If there is more than one clause in each set that remains, then there are multiple assignments that are the unions of the clauses of each set respectively.

## Description of 3SAT

SAT stands for satisfiability. It is a propositional problem of determining if there is any assignment of truth-values for the variables in such a way that the whole formula is true. The formula is prepositional calculus in conjunctive normal form (CNF), in which each clause contains 3 variables in disjunctive form. The following is an instance, where ~ indicates NOT:

$$F = (x_1 \vee x_2 \vee x_3) \wedge (x_1 \vee x_2 \vee \sim x_3) \wedge (x_1 \vee x_2 \vee x_4) \wedge (x_1 \vee \sim x_4 \vee \sim x_5) \wedge (x_2 \vee \sim x_3 \vee x_5)$$

Formula $F$ has $m$ clauses (denoted by parentheses) in conjunctive form, $n$ variables ($x_1$, $x_2$, $x_3$… $x_n$) in disjunctive form of three variables. To solve an instance of this decision problem we have to determine whether there is a truth-value (TRUE or FALSE) we can assign to each of the variables such that the entire expression is TRUE. If there were no such an assignment, the answer would be FALSE.

From KRA aspect, all languages are organized in the form of member-class iterative set relations as knowledge between two "mirrored" languages [Han 10]. An instance of 3SAT is a set of specific knowledge. That is, each 3SAT problem contains $m$ disjunctive clauses, and each disjunctive clause can be represented by a set of eight conjunctive clauses (COVA) including one complement. For example, disjunctive clause $(x_1 \vee x_2 \vee \sim x_3)$ can be presented by a set of eight conjunctive clauses $\{(x_1 \wedge x_2 \wedge x_3), (x_1 \wedge x_2 \wedge \sim x_3), (x_1 \wedge \sim x_2 \wedge x_3), (x_1 \wedge \sim x_2 \wedge \sim x_3), (\sim x_1 \wedge x_2 \wedge x_3), (\sim x_1 \wedge x_2 \wedge \sim x_3), (\sim x_1 \wedge \sim x_2 \wedge x_3), (\sim x_1 \wedge \sim x_2 \wedge \sim x_3)\}$ in which a 3-variable-clause $(\sim x_1 \wedge \sim x_2 \wedge x_3)$ is a complement of $(x_1 \vee x_2 \vee \sim x_3)$.

---

[*] The principle of COVA indicates that eight 3-variable conjunctive clauses are the basic units for numerical and logical operations.

In the knowledge (iterative set) structure, each 3-variable-clause $C_3$ contains three 2-variable-clause $C_2$ as its members. For example, clause $(x_1 \wedge x_2 \wedge x_3)$ contains three 2-variable-clauses $\{(x_1 \wedge x_2), (x_1 \wedge x_3), (x_2 \wedge x_3)\}$ as its members. Each 2-variable-clause $C_2$ may belong to multiple 3-variable-clauses $C_3$.

Each 2-variable-clause $C_2$ contains two variables that represent a value of either 0 or 1 as its members. Each 4-variable-clause $C_4$ contains four 3-variable-clause $C_3$ as its members. Each 3-variable-clause $C_3$ may belong to multiple 4-variable-clauses $C_4$.

## Descriptions of the KRA Procedure on 3SAT

Different from other 3SAT algorithms, KRA does not search assignment directly. It recognizes the complements as rejections at each level of the set through relation recognition iteratively. That is, KRA recognizes which conjunctive 3-variable-clause is not satisfiable through set operations. If all the eight clauses of any set of 3-variable-calsues (a set of COVA) are rejected, then there is not assignment for the formula. If there is at least one clause in each set that remains, then there is at least one assignment that is the union of clauses of each set. If there are more than one clause in each set that remain, then there are multiple assignments that are the unions of the clauses of each set respectively. The algorithm procedure can be described as follows.

1. Convert the disjunctive clauses to conjunctive complement clauses for rejection. For example, $(x_1 \vee x_2 \vee \sim x_3)$ can be converted as $(\sim x_1 \wedge \sim x_2 \wedge x_3)$, which is a rejection.
2. *Recognize* complements through member-class relations $R=\{\Sigma \in \Sigma_{|2|}^* \in \Sigma_{|3|}^* \in \Sigma_{|4|}^*\}$ deductively and reductively. Rules of complement recognition are described in the following section.
3. 3SAT formula is constructed with a binary alphabet $\Delta =\{0, 1\}$ such that $\Delta^* \wedge \{\text{ACCEPT, REJECT}\} = \phi$. The algorithm recognizes only the REJECTIONS of the domain. The ACCEPTION is the difference of the domain.
4. Range R = domain D = $8n^3$ = $\{\Sigma \in \Sigma_{|2|}^* \in \Sigma_{|3|}^* \in \Sigma_{|4|}^*\}$ = {ACCEPT, REJECT} (the space is eight octants or COVA of the three-dimensional coordinate system).
5. Encoding mapping E: D $\rightarrow \{\Sigma \in \Sigma_{|2|}^* \in \Sigma_{|3|}^* \in \Sigma_{|4|}^*\}$
6. Transition $\tau$: $=\{\Sigma \in \Sigma_{|2|}^* \in \Sigma_{|3|}^* \in \Sigma_{|4|}^*\} \rightarrow \{\Sigma \in \Sigma_{|2|}^* \in \Sigma_{|3|}^* \in \Sigma_{|4|}^*\} \cup \{\text{ACCEPT, REJECT}\}$.

## Rules of rejection (complement recognition)

3SAT complements can be recognized at four levels through rejection rules:

**1-variable complement recognition**

1) If one pair of variables is complement, one pair of variables is an intersection between two rejected 2-variable-clauses such as (a, x) and (a ~x), where (x) and (~x) are complements, (a) is the intersection, then the variable value (a) is rejected. It is denoted by (2-2CI), where 2-2 indicates the variable numbers of the related clauses, C indicates complement, I indicates intersection, and D indicates difference.

**2-variable complement recognition**

1) If one pair of variables is complement, two pairs of variables are intersections between two rejected 3-variable-clauses such as (a, b, x) and (a, b, ~x), where (x) and (~x) are complements, (a, b) are the intersections, then the 2-variable-clause (a, b) is rejected. It is denoted by (3-3CII).

**3-variable complement recognition**

1) If any 3-variable-clause contains any rejected value of a variable, such as (a, b, c) contains (a) and (a) is a rejected value, then clause (a, b, c) is rejected. It is denoted by (1-3I).

2) If any 3-variable-clause contains any rejected 2-variable-clause, such as (a, b, c) contains (a, b) and (a, b) is a rejected 2-variable-clause, then clause (a, b, c) is rejected. It is denoted by (2-3II).

3) If one pair of variables is complement, the rest of variables are differences between a rejected 2-variable-clause and a rejected 3-variable-clause such as (a, x) an (b, c, ~x), where (x) and (~x) are complements, (a), (b), and (c) are differences, then clause (a, b, c) is rejected. It is denoted by (2-3CDD).

4) If one pair of variables is complement, one pair of variables is an intersection, and two other variables are differences between a rejected 2-variable-clause and a rejected 4-variable-clause such as (a, x) and (a, b, c, ~x), where (x) and (~x) are complements, (a) is the intersection, (b) and (c) are the differences, then clause (a, b, c) is rejected. It is denoted by (2-4CIDD).

5) If one pair of variables is complement, one pair of variables is an intersection, and two other variables are differences between two rejected 3-variable-clauses such as (a, b, x) and (a, c, ~x), where (x) and (~x) are complements, (a) is the intersection, (b) and (c) are the differences, then clause (a, b, c) is rejected. It is denoted by (3-3CID).

6) If one pair of variables is complement, two pairs of variables are intersections, one variable is a difference between a rejected 3-variable-clause and a rejected 4-variable-clause such as (a, b, x) and (a, b, c, ~x), where (x) and (~x) are complements, (a, b) are the intersections, and (c) is the difference, then clause (a, b, c) is rejected. It is denoted by (3-4CIID).

7) If one pair of variables is complement, three pairs of variables are intersections between two rejected 4-variable-clauses such as (a, b, c, x) and (a, b, c, ~x), where (x) and (~x) are complements, (a), (b), and (c) are the intersections, then clause (a, b, c) is rejected. It is denoted by (4-4CIII).

**4-variable complement recognition**

1) If one pair of variables is complement, two pairs of variables are differences between two rejected 3-variable-clauses such as (a, b, x) and (c, d, ~x), where (x) and (~x) are the complements, (a), (b), (c) and (d) are differences, then the 4-variable clause (a, b, c, d) is rejected. It is denoted by (3-3CDD).

## General structure of the algorithm

```
KRA_3SAT()                                              //KRA rejects complements
    if every clause of a set is rejected then
        return false and no assignment
    else if no new rejected complement then
        return true and assignment                       //union of accepted clauses
    else
        New_Complement_Rejection (Σ ∈ Σ*_{|2|} ∈ Σ*_{|3|} ∈ Σ*_{|4|})    //set operations
        go to KRA_3SAT()                                 //exhaust search
    end if
```

## Postscript

Three hundred years ago Leibniz wrote with joy that he discovered the meaning of COVA: "The Chinese lost the meaning of the COVA or Lineations of Fuxi, perhaps more than a thousand years ago, and they have written commentaries on the subject in which they have sought I know not what far out meanings, so that their true explanation now has to come from Europeans. Here is how: It was scarcely more than two years ago that I sent to Reverend Father Bouvet, the celebrated French Jesuit who lives in Peking, my method of counting by 0 and 1, and nothing more was required to make him recognize that this was the key to the figures of Fuxi. Writing to me on 14 November 1701, he sent me this philosophical prince's grand figure, which goes up to 64, and leaves no further room to doubt the truth of our interpretation, such that it can be said that this Father has deciphered the enigma of Fuxi, with the help of what I had communicated to him. And as these figures are perhaps the most ancient monument of science which exists in the world, this restitution of their meaning, after such a great interval of time, will seem all the more curious." [Leibniz 1703]

What Leibniz discovered was the numerical (binary) meaning of COVA, which enabled **machine computation**. There was another meaning of logic, the law of **reducibility**, which was lost as well. **Reducibility** is the central important philosophical and mathematical notion behind 3SAT and other NP-complete problems [Coo00, Karp72, Wigderson 09]. The mechanism of KRA is based on such a reducibility of COVA, which enables **machine recognition**.

## Acknowledgement

Our thanks to Luo Fangbin, Zhou Guoqing, Fan Yaxi, Zhu Guoping, Yan Weimin, Che Zhibin, Sui Jianfeng, Lin Guang, Steve Chien, Samuel Eaves, Jose Cid, and Ming Shao for their technical and engineering contribution in this teamwork.